\documentclass[prl,reprint]{revtex4-1}

\usepackage{amssymb}
\usepackage{amsmath}
\usepackage{latexsym}
\usepackage{graphicx}
\usepackage{psfrag}
\usepackage{epstopdf}
\begin{document}

\title{The One-Dimensional KPZ Equation: an Exact Solution
and its Universality} 
\date{today}
\author{Tomohiro Sasamoto}
\affiliation{Department of Mathematics and Informatics, 
 Chiba University, 1-33 Yayoi-cho, Inage, Chiba 263-8522, Japan}
\email{sasamoto@math.s.chiba-u.ac.jp}
\author{Herbert Spohn}  
\affiliation{Zentrum Mathematik and Physik Department, TU M\"unchen,
 D-85747 Garching, Germany}
 \email{spohn@ma.tum.de}
\begin{abstract}
 We report on the first exact solution of the
KPZ equation in one dimension, with an initial condition which physically corresponds
to the motion of a macroscopically curved height profile. 
The solution provides a determinantal formula for the
probability distribution function of the height $h(x,t)$ for all
$t>0$. In particular, we show that for large $t$, on the scale
$t^{1/3}$, the statistics is given by the Tracy-Widom
distribution, known already from the theory of GUE random matrices.
Our solution confirms that the KPZ equation describes the interface
motion in the regime of weak driving force. Within this regime the KPZ equation 
details how the long time asymptotics is approached.
\end{abstract}
\maketitle
The motion of interfaces persists as a fascinating topic of statistical mechanics. One particular, intensely studied case
are nonequilibrium growth processes,
which are governed by \textit{local} rules. In the seminal work \cite{KPZ} Kardar, Parisi, and Zhang proposed a 
model equation, now called KPZ equation, to investigate the dynamic scaling of such
growing interfaces.  KPZ argued that a growing interface has a statistically self-similar structure with universal scaling exponents for the interface width and for the transverse correlation length. In particular, they predicted that a one-dimensional interface has fluctuations which grow as $t^{1/3}$, in contrast to the $t^{1/4}$ statistical broadening of an equilibrium interface. Details of the KPZ scenario have been investigated through Monte Carlo simulations of simplified stochastic growth models,  like Eden growth, random deposition, and polynuclear growth. Universal scaling exponents were confirmed, supporting that all such growth models constitute the KPZ universality class. On the theoretical side a variety of techniques have been devised, useful also in other areas of nonequilibrium statistical mechanics, as dynamic and functional renormalization group,  mode-coupling theory, exact solutions, and more. For the development up to 1995 we refer  to the book \cite{BS} and the reviews \cite{HZ,K}.
 However, so far the KPZ equation itself has resisted analytical handling. We will report here on the first exact solution. Physically it describes cluster growth in a thin film. The solution also provides a better understanding of the universality of the KPZ equation.

We will
be concerned with the one-dimensional KPZ equation only. The interface
location is then described by a height function $h(x,t)$ depending
on time $t$ and location $x$ on the real line. The KPZ
equation reads
\begin{equation}\label{1}
\frac{\partial}{\partial t}h= \tfrac{1}{2}
\lambda\big(\frac{\partial}{\partial x}h\big)^2
+\nu\frac{\partial^2}{\partial x^2}h +\sqrt{D}\eta\,.
\end{equation}
The first term is the slope dependent growth velocity of strength
$\lambda$, $\lambda>0$ for convenience. The Laplacian smoothens the interface with
diffusivity $\nu>0$. $\eta$ is normalized space-time white noise
with correlator $\langle\eta(x,t)\eta(x',t')\rangle=\delta(x-x')
\delta(t-t')$ and $\sqrt{D}$ is the noise strength. $\eta$ models
the random nucleations at the interface. We will choose initial conditions such that the height profile 
remains curved on the macroscopic scale.

 In the breakthrough contribution  \cite{J} Johansson firstly succeeded to compute a universal probability distribution function (pdf), see also the related work 
 \cite{BDJ,PS}.  He studied a discrete growth model, known as single step, with wedge initial profile.  Most surprisingly, he discovered that the pdf for the random amplitude of the height is 
 the Tracy-Widom distribution
\cite{TW1}, first obtained in the context of the large $N$
statistics of the largest eigenvalue of a GUE random matrix
\cite{For}. The Tracy-Widom pdf, $\rho_\mathrm{TW}(s)$, is defined through the determinant of a
symmetric operator acting on  functions on the real line, where, in principle, the determinant is defined by the 
product of the eigenvalues. More explicitly, we denote 
 by $P_s$ the projection operator onto the interval $[s,\infty)$ and by $K_\mathrm{Ai}$ the Airy kernel
 \begin{equation}\label{5a}
K_{\mathrm{Ai}}(x,y)=\int^\infty_0 \mathrm{d}w
\mathrm{Ai}(x+w) \mathrm{Ai}(y+w)
\end{equation}
with Ai the standard Airy
function. Then
\begin{equation}\label{5}
\rho_\mathrm{TW}(s)= \frac{d}{ds}\det (1-P_s K_\mathrm{Ai}P_s)\,.
\end{equation}
One can show that $\mathrm{tr}[P_s K_\mathrm{Ai}P_s] < \infty$ \cite{TW1} and hence the infinite product makes sense.

In a spectacular recent experiment  on electroconvection \cite{T}
the Tracy-Widom statistics is verified down to a scale of height samples
with probability $10^{-4}$. In this experiment  
a thin film of liquid crystal is electrically driven to a turbulent phase, 
the unstable DSM1 phase. 
One then plants through a laser pulse a sharply localized seed of topological-defect turbulence, the  
stable DSM2 phase. It grows isotropically over a time window of 30 sec to the maximal size of 1.6 cm in diameter.
The experiment also investigates how the Tracy-Widom distribution is approached for long times. One observes that the cumulants 2, 3, and 4 have already reached their Tracy-Widom value, while the average still slowly decays 
as $t^{-1/3}$ towards its stationary value,
an observation so far without theoretical explanation.

\textit{Exact solution}.---The curved height profile can locally be approximated by a parabola. In idealization,
we therefore choose our initial conditions such that the average height is exactly parabolic, which can be achieved
through the initial sharp wedge 
\begin{equation}\label{2}
h(x,0)=-|x|/\delta\,,\quad \delta\ll 1\,.
\end{equation}
Since for short times the nonlinear term dominates,
the solution $h(x,t)$ spreads
rapidly into the parabolic profile,
$h(x,t) \simeq -x^2/2\lambda t$ for $|x|\leq \lambda t/\delta$ and 
$h(x,t) \simeq (\lambda/2\delta^2)t-|x|/\delta$ for $|x|\geq\lambda t/\delta.$ Because of the noise term in (\ref{1})
this profile has superimposed random fluctuations whose amplitude grows as $t^{1/3}$. 
A typical realization is shown in Fig. 1(a).


\begin{figure}
 \begin{picture}(250,100)(,)
  \put(8,95){(a)}
  \put(10,10){\includegraphics[scale=0.6]{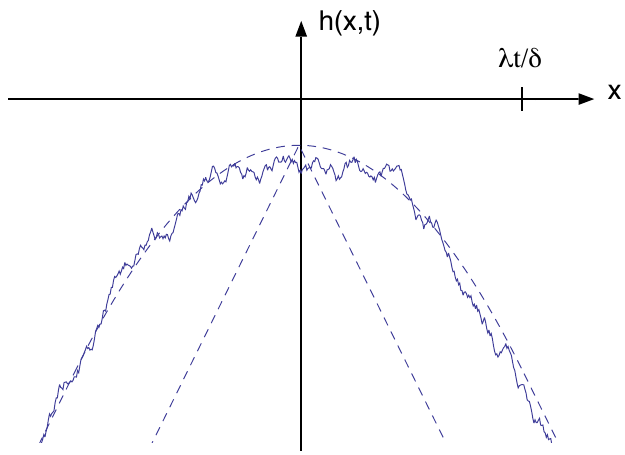}}  
  \put(138,95){(b)}
  \put(145,0){\includegraphics[scale=0.4]{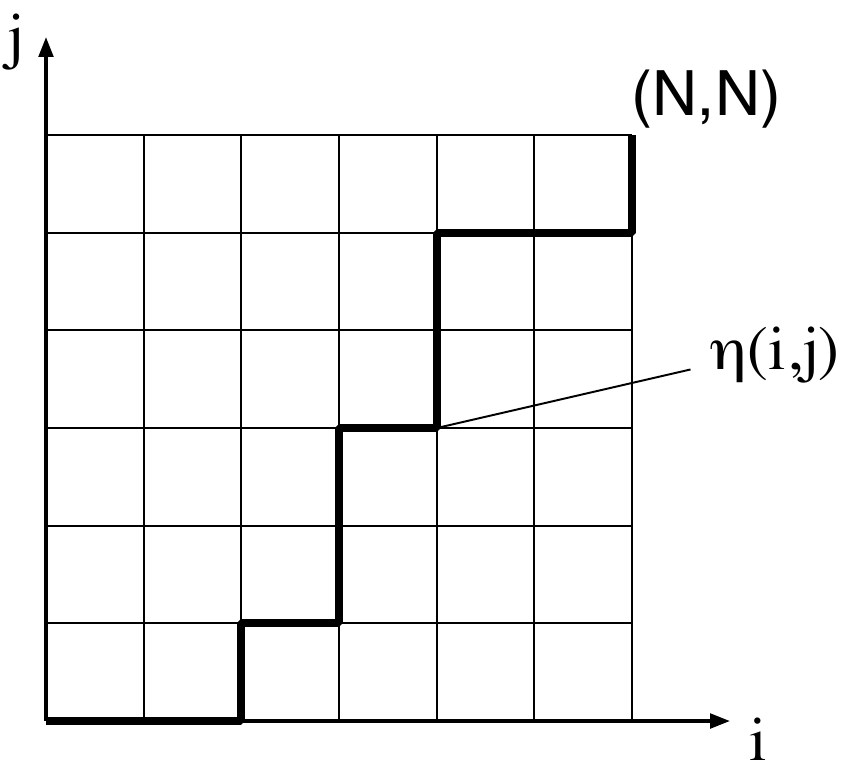}}
 \end{picture}
\caption{(a) A typical realization of the droplet height function.
(b) A directed polymer configuration. 
}
\end{figure}

We consider the height
statistics at one prescribed point $x$. Then, for every $t>0$,
\begin{equation}\label{6}
(\lambda/2\nu) h(x,t) 
= -x^2/4\nu t-\tfrac{1}{12}\gamma_{t}^3 +
2\log\alpha + \gamma_{t}\xi_{t}\,,
\end{equation}
where
\begin{equation}\label{7a}
\gamma_t =  ( \alpha^4\nu t)^{1/3}\,,\quad
\alpha=(2\nu)^{-3/2}\lambda D^{1/2}\,.
\end{equation}
The first three terms of (\ref{6}) are deterministic, in particular one notes 
the inverted parabola already mentioned.  The logarithmic term reflects the 
scale invariance of the initial sharp wedge. All information on the fluctuations of $h(x,t)$ is
encoded in the random amplitude $\xi_t$, which extends the Tracy-Widom amplitude,
$\xi_\mathrm{TW}$, to
finite time $t$.
The prefactor $\gamma_t$ confirms that the fluctuations grow as $t^{1/3}$. 
The pdf of $\xi_t$ is given by
\begin{eqnarray}\label{7}
&&\hspace{-10pt}\rho_t (s)= \int^\infty_{-\infty} \mathrm{d}u \,\gamma_t
\mathrm{e}^{\gamma_t(s-u)}\exp \big[-\mathrm{e}^{\gamma_t(s-u)}\big]\nonumber\\
&&\hspace{-10pt}\times\big(\det (1-P_u
(B_t-P_\mathrm{Ai})P_u) -\det(1-P_u
B_tP_u)\big)\,.
\end{eqnarray}
Here $P_\mathrm{Ai}$ has the integral kernel $\mathrm{Ai}(x)\mathrm{Ai}(y)$, which can be viewed
as a one-dimensional unnormalized projection, and $B_t$ has the kernel
\begin{equation}\label{8}
B_t(x,y)= \int^\infty_{-\infty}
\mathrm{d}w
(1- \mathrm{e}^{-\gamma_t w})^{-1}
\mathrm{Ai}(x+w)
\mathrm{Ai}(y+w)\,.
\end{equation}

$\xi_t$ is independent of $x$ and depends on $t $ only through the
dimensionless parameter $\gamma_t$.
In the
limit $t\to\infty$ the Gumbel density in (\ref{7}) tends
to $\delta(s-u)$ and $B_t$ tends to $B_\infty=K_\mathrm{Ai}$. 
The difference of determinants in (\ref{7}) is then precisely $\rho_\mathrm{TW}(s)$. 
Hence  $\xi_t $ tends to 
$\xi_\mathrm{TW}$ as $t \to \infty$.  In case of a macroscopically curved profile, 
it is expected that for any growth process in the 
KPZ class the pdf of the height fluctuations 
is Tracy-Widom in the long time limit.
We have thus established that the KPZ equation is in the KPZ universality class as regards to the 
pdf of the height fluctuations.

Our exact solution also provides an explanation for the observed $t^{-1/3}$ 
relaxation to stationarity. The difference of determinants in (\ref{7}) is of order $t^{-4/3}$ \cite{SS2}. 
Hence the leading correction to
$\rho_\mathrm{TW}(s)$ is the shift due to the non-zero mean of the Gumbel distribution 
which is of order $t^{-1/3}$. 

To represent the solution to (\ref{1}) in a more tractable form we introduce the 
Cole-Hopf  transformation as
\begin{equation}\label{9}
Z(x,t)=\exp[(\lambda/2\nu) h(x,t)]\,.
\end{equation}
At the expense of multiplicative noise, $Z(x,t)$ then satisfies the linear equation
\begin{equation}\label{10}
\frac{\partial}{\partial t}Z=\nu \frac{\partial^2}{\partial
x^2}Z+(\lambda\sqrt{D}/2\nu)\eta Z\,.
\end{equation}
with the normalized initial condition $Z(x,0) = \lim_{\delta \to 0} ((2\nu/\lambda)2 \delta)^{-1}
\exp[- (\lambda/2\nu)|x|/\delta] = \delta(x)$.  (\ref{10}) 
is solved through the Feynman path integral
\begin{equation}\label{11}
Z(x,t)=\mathbb{E}_0\Big(\exp\Big[\alpha
\int^{2\nu t}_0 \mathrm{d}s
\eta(b(s),s)
\Big]\delta(b(2\nu t)-x)\Big)\,.
\end{equation}
Here $ \mathbb{E}_0(\cdot)$ is the Wiener integral over all paths of an auxiliary
Brownian motion $b(t)$, starting at $0$ and with variance $t$.
In principle, the solution of the KPZ equation with sharp wedge
initial condition is defined by
\begin{equation}\label{13}
h(x,t)=(2\nu/\lambda)\log Z(x,t)\,.
\end{equation}
However, as written $\langle Z(x,t) \rangle = \infty$ because of ultraviolet divergencies. 
As in quantum field theory one thus has to introduce a suitable cutoff to be 
removed through a renormalization scheme. We briefly describe four distinct 
variants, all providing physical insight to the interpretation of the KPZ equation. 
But only the last variant carries the computational power to obtain the exact solution.

\textit{(i) Colored noise}.---The expression (\ref{11}) is only formal, since the white 
noise $\eta$ is very rough and hence the action in (\ref{11}) relies on an undefined 
integral. One way to improve the situation is to substitute in the KPZ equation (\ref{1}) 
the noise $\eta(x,t)$ by a noise $\eta_{\kappa}(x,t)$, which is colored in $x$ with 
width $\kappa^{-1}$ but still white in $t$, such that $\eta_{\kappa} \to \eta$ 
as $\kappa \to \infty$. This modification induces the uniform translation of the 
height profile by $v_\kappa t$ with $v_\kappa  \simeq \kappa$. Going to the 
moving frame of reference, the limit $\kappa \to \infty$ of the height profile is 
well defined \cite{ACQ,BG} and yields the proper interpretation of (\ref{11}).

\textit{(ii) Directed polymer in a random potential}.---The integral in (\ref{11}) is discretized. 
Then the Brownian motion path $b(t)$ is replaced by  a directed polymer $\omega$, which 
for convenience is placed on the two-dimensional lattice $\mathbb{Z}^2$. It starts at $0$, 
makes only up/right moves, and ends at the lattice site $(tN,tN)$, 
see Fig. 1(b). Independently for each site $(i,j)\in \mathbb{Z}^2$, 
there is a unit Gaussian random potential $\eta(i,j)$.
 The energy of the directed
polymer is $E(\omega)=\sum\eta(i,j)$, where the sum is over the potentials along the
polymer $\omega$. The discrete approximation to (\ref{11}) reads then 
\begin{equation}\label{14}
Z_{tN}= \sum_{\omega:(0,0)\leadsto (tN,tN)} \exp[-\beta E(\omega)]\,.
\end{equation}
The partition function  is point-to-point, since both endpoints of the polymer are fixed. In the limit $N\to \infty$, $\beta \to  0$ with $\beta^4N$ fixed,
$Z_{tN} \to Z(0,t)$
and $Z(0,t)$ as defined in (i) \cite{AKQ}. From this perspective the KPZ equation is the weak noise limit of the directed polymer, see \cite{CLR} for details.

\textit{(iii) Attractive $\delta$-Bose gas} \cite{K2}.---The $n$-th moment, $\langle Z(x,t)^n \rangle$, of the partition function (\ref{11}) can be expressed 
through the propagator of $n$ quantum particles on the line interacting through an attractive $\delta$-potential. All $n$ particles start at $0$ and propagate to $x$ at time $t$. Working out  
$\langle Z^n \rangle$ yields the potential
$-\tfrac{1}{2} \alpha^2 \sum_{i,j=1}^{n} \delta (x_i - x_j)$.
Thus the renormalization corresponds to merely removing the self-energy through normal ordering. Thereby one arrives at the same random partition function as constructed in (i) and  (ii).

\textit{(iv) Single step growth model}.---This is a stochastic evolution model for the integer valued height
function $h(j,t)$, $j\in\mathbb{Z}$, satisfying the single step
constraint
\begin{equation}\label{8a}
|h(j+1,t)-h(j,t)|=1\,.
\end{equation}
In our case the initial height profile is the wedge $h(j,0)=-|j|$. A random sequential update rule is
used: Independently the height at a local minimum is increased
by 2 with rate $p$ and at a local maximum decreased by 2 with rate
$q$, $p+q=1$ to set the time scale, and $q>p$ in our case 
corresponding to $\lambda > 0$. The
height differences are then governed by the partially asymmetric simple
exclusion process (PASEP). This is a stochastic particle system on
$\mathbb{Z}$, where there is at most one particle per site. Particles
jump with rate $p$ to the right and rate $q$ to the left under the
constraint of the exclusion rule. The initial wedge corresponds
to the 0\,-1 step initial particle configuration, for which all sites to left of
the origin are empty and to the right of the origin are filled. In
\cite{J} the Tracy-Widom fluctuations are proved in case $p=0$,
$q=1$ (the TASEP), including a discrete time parallel random update rule. The
intricate extension to $0<p<q<1$ has been accomplished by Tracy and
Widom \cite{TW2,TW3}. Also available is the corresponding result for
the PNG droplet \cite{BDJ,PS1,BFS}. 

The Cole-Hopf solution (\ref{11}) arises at the crossover scale to weak asymmetry
(WASEP). More specifically,  
 one assumes
$q = \tfrac{1}{2} + \sqrt{\varepsilon}\beta$, $\beta > 0$, $\varepsilon>0$, and $\varepsilon\ll
1$. The correspondingly adjusted time scale is order $\varepsilon^{-2}$ and space scale is order $\varepsilon^{-1}$,
while the height is of order $\varepsilon^{-1/2}$. If $h^\varepsilon (j,t)$ denotes the WASEP random height profile, then the single step partition function is defined through 
\begin{equation}\label{8b}
Z^{\varepsilon}_\mathrm{step}(j,t) = \exp[ (\log(p/q))h^\varepsilon(j,t)]\,.
\end{equation}
As shown in \cite{BG,ACQ}, the correctly centered $Z^{\varepsilon}_\mathrm{step}(\varepsilon^{-1}x,\varepsilon^{-2}t)$
converges in the limit $\varepsilon \to  0$ to $Z(x,t)$, constructed already before in (i) - (iii). 

We are now in a position to  indicate, rather roughly, how our exact solution  is obtained. A more detailed exposition can be found in \cite{SS1,SS2}. 
For the 0\,-1 step initial particle configuration, Tracy
and Widom  \cite{TW2} recently provided a contour integration
formula for the probability distribution of the position of the $m$-th particle at
time $t$ valid for arbitrary $q$.  In \cite{SS1} we use their formula as starting point
and study the scaling limit of the PASEP as above, namely asymmetry
$\sqrt{\varepsilon}\beta$, space $\mathcal{O}(\varepsilon^{-1})$,
and time $\mathcal{O}(\varepsilon^{-2})$, to arrive at a pdf given as an integral
over the difference of two determinants as in (\ref{7}), including a
deterministic shift of order $ \varepsilon^{-1}$ and a
subleading $\log \varepsilon$ correction. As discussed in the
companion paper \cite{SS2}, in addition one has to analyze the
average $\langle Z^{\varepsilon}_\mathrm{step}(j,t) \rangle$ for the WASEP, thereby
to determine the appropriate centering. One finds terms of order  $ \varepsilon^{-1}$ and $\log
\varepsilon$, which are precisely cancelled by the  corresponding
terms appearing in the WASEP scaling limit. Combining both results,
one arrives at (\ref{6}) - (\ref{8}).
The KPZ parameters are fixed through the WASEP as $2\nu = 1$ 
and $\beta = \lambda$. At $x=0$ the WASEP average density is $1/2$ and hence the noise strength $D = 1/4$. 
By varying $x$ one can tune the noise strength to $\rho(1 - \rho)$, where $\rho$ is the local average density. In principle it suffices to
consider $x=0$ and to deduce the parameter dependence from (\ref{11}).

For a more detailed information on the pdf $\rho_t(s)$ one has to rely on a numerical evaluation of 
(\ref{7}), for which purpose it is 
 convenient to regard, \textit{e.g.}, 
$P_u B_t P_u$ as a large matrix. 
One then evaluates  $B_t(x,y)$ 
at a judiciously chosen set of grid points $x_j$, $j = 1, \dots, n$, $u \leq x_j$, 
with $n$ of order 60 to 120 and thereby defines the $n\times n$ 
matrix $B_{ij} =  B_t(x_i,x_j)$, $i,j=1,\dots,n$, 
for which $\mathrm{det}(1-B)$ is obtained by a standard routine. 
Since the kernel of (\ref{8}) is a smooth function, the computation 
is fast and the errors are small \cite{Bo}. 
In Fig. 2(a) we display the pdf $\rho_t(s)$ for $\gamma_t = 2,5,10$ obtained by the method described.
For even smaller values of $\gamma_t$ the computation
becomes slow because of strong oscillations of the kernel $B_t(x,y)$
resulting from the singularity at $w=0$ of the defining
integral (\ref{8}).


\begin{figure}
 \begin{picture}(250,80)(,)
  \put(20,70){(a)}
  \put(5,0){\includegraphics[scale=0.37]{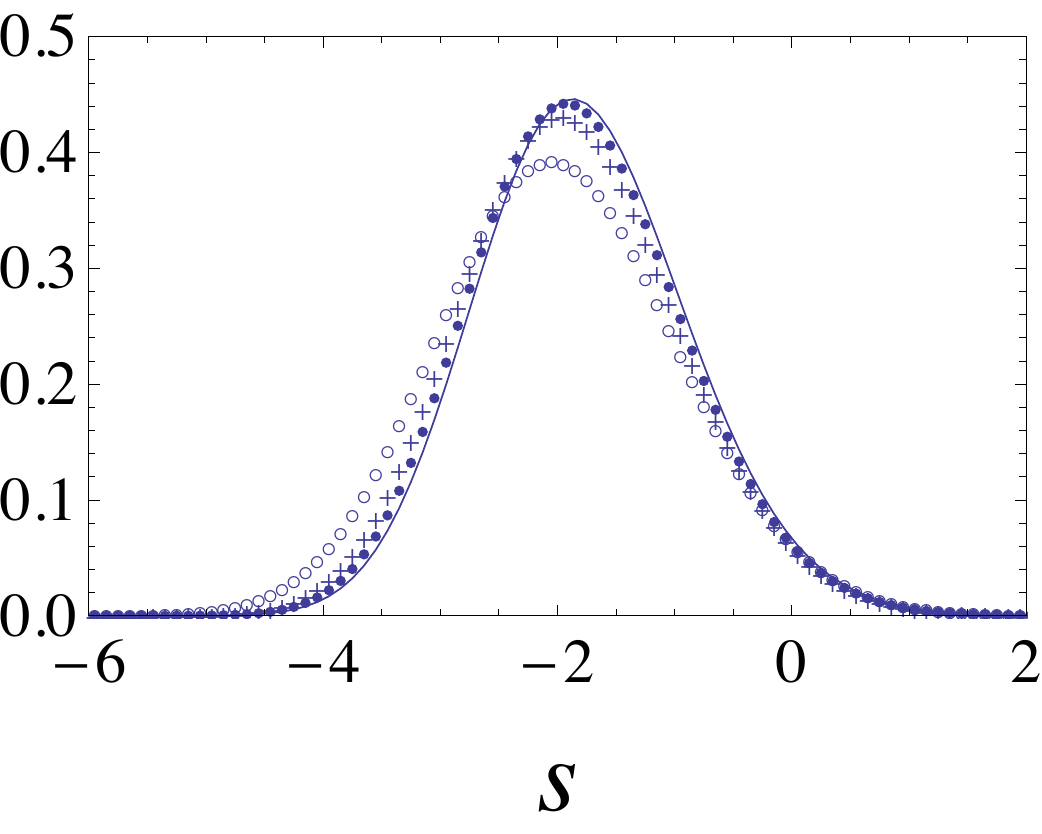}}  
  \put(140,70){(b)}
  \put(120,0){\includegraphics[scale=0.37]{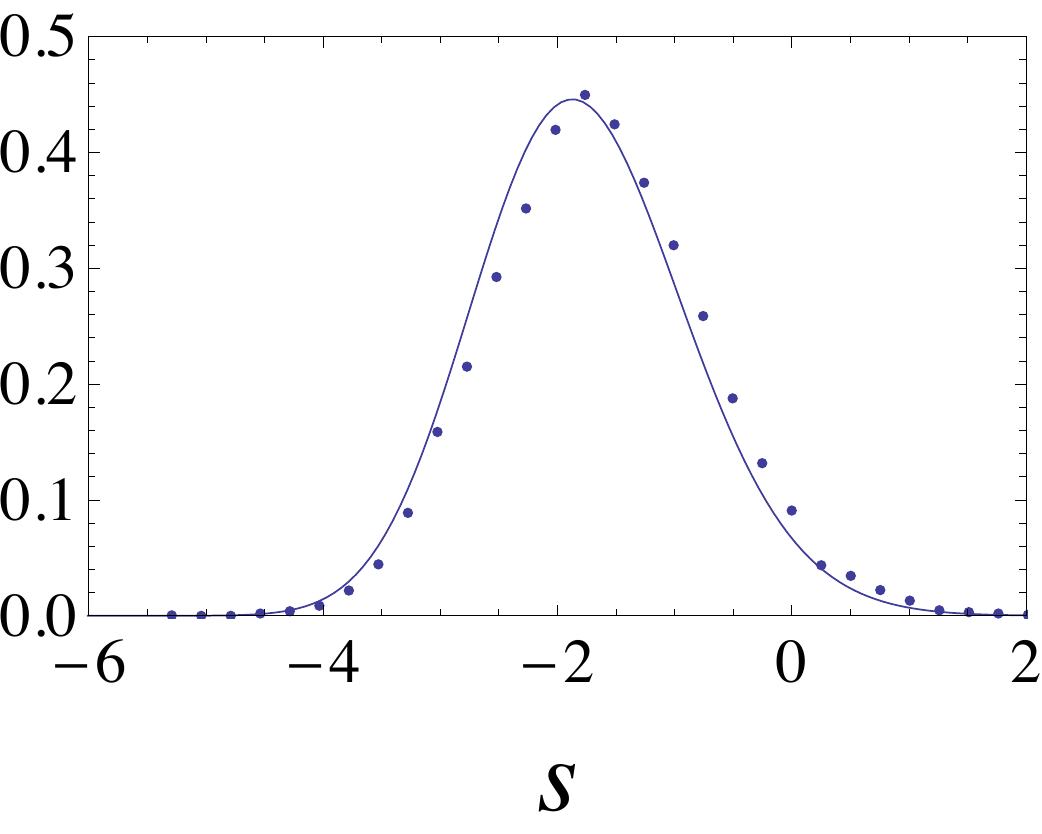}}
 \end{picture}
\caption{(a) The pdf defined by (\ref{7}) for $\gamma_t=2(-),5(+),10(\cdot)$ and the Tracy-Widom 
pdf defined by (\ref{5}) (full curve).
(b) The TASEP height statistics at $10^3$ MC steps and the Tracy-Widom pdf.}
\end{figure}

\textit{Universal properties}.---The exact solution has been obtained through a limit of vanishing
asymmetry. More generally, one could consider some growth model in the KPZ class with a tunable asymmetry, denoted by $\beta$. Physically the asymmetry
is possibly small, but fixed. Time is measured in Monte Carlo time steps.
For times of order $\beta^{-2}$ the nonlinearity plays no role, yet,
and the height fluctuations are approximately
Gaussian. On the time scale $\beta^{-4}$, one crosses
over to the KPZ solution (\ref{6}) with non-Gaussian statistics,
which asymptotically is well approximated by the Tracy-Widom
distribution. Of course, a clean separation of time scales is
expected only for $\beta \ll 1$. 

To have a test case we performed MC
simulations for the PASEP with strongest possible asymmetry, namely
$q=1$, $p=0$ (the TASEP). For the 0\,-1 step initial condition we sampled
the height at the origin, $h_{q=1}(0,t)$. The scale linear in $t$ is given by
$-(q-p)t/2$. For $t=10^3$ MC steps the distribution of
$h_{q=1}(0,t)$ is concentrated approximately in an interval of 60 lattice units. 
As displayed in  Fig. 2(b), even on the discrete level
the Tracy-Widom distribution is still an accurate approximation,
which becomes almost indistinguishable upon shifting it by $0.13$ units to the right.
The $t^{-1/3}$ shift to
the right was noted before numerically from the recursion relations of the PNG droplet \cite{P} and is also
observed in the experiment \cite{T}.
According to the exact solution, for large $t$,
$\rho_t$ is shifted by $\gamma_t^{-1}(-0.577 + 2 \log\alpha)$ relative to
$\rho_\mathrm{TW}$, where $-0.577 $ is the mean of the Gumbel
density. Since $\alpha$ is proportional to $\lambda$, this suggests that
the prefactor of the $t^{-1/3}$ shift for the PASEP should undergo a sign 
change. Indeed, running the MC simulation for smaller values of
$q$, one finds that  $\rho_t$ is shifted to the left side
of $\rho_\mathrm{TW}$ for $q\lesssim 0.78$. The approach as $t^{-1/3}$
seems to be universal and is expected to hold for any growth model in the KPZ class. 
However the precise first order correction 
to $\rho_{\mathrm{TW}}(s)$ will depend on the particular model. 

\textit{Conclusions}.---We have computed the exact probability distribution function for the 
height $h(x,t)$ of the KPZ equation with narrow wedge initial profile. 
The long time limit is the Tracy-Widom pdf, in accordance
with the results for discrete growth models. But in addition, based on our exact solution, 
we now better understand how the long time asymptotic is approached. For growth models with tunable asymmetry the KPZ equation is an accurate approximation for weak asymmetry.

\textit{Note}.---After submitting our letter,  three distinct groups \cite{ACQ,CLR,D}
posted independent but related work on the arXiv.
 
We are grateful to Michael Pr\"{a}hofer
for many illuminating discussions. H.~S. thanks Jeremy
Quastel for emphasizing the importance of the crossover WASEP.
This work is supported by a DFG grant. In addition T.S. acknowledges the support from
KAKENHI (9740044) and H.S. from Math-for-Industry
of Kyushu University.


\end{document}